\documentclass[twocolumn]{aastex7}

\renewcommand{\baselinestretch}{1.1}
\usepackage{textcomp}
\usepackage{gensymb}
\usepackage{epstopdf}
\usepackage{amsmath}
\usepackage{graphicx}
\usepackage{hyperref}
\usepackage{multirow}
\usepackage[nameinlink]{cleveref}
\usepackage[multiple]{footmisc}
\crefname{paragraph}{\S}{\S\S} 
\newcommand{\MORIA}{\texttt{MORIA} }
\shorttitle{MORIA} 
\shortauthors{Bhadra et al.}

\begin{document}
\title{\textbf{You Shall Not Pass (Without Modeling): High-Resolution Analysis of KMT-2019-BLG-0253 using MORIA}}

\author[0009-0002-6097-9030]{T. Dex Bhadra}
\affiliation{Department of Astronomy, University of Maryland, College Park, MD 20742, USA}
\email[show]{tmbhadra@umd.edu}

\author[0000-0002-5029-3257]{Sean K. Terry}
\affiliation{Department of Astronomy, University of Maryland, College Park, MD 20742, USA}
\affiliation{Code 667, NASA Goddard Space Flight Center, Greenbelt, MD 20771, USA}
\email{skterry@umd.edu}

\author[0000-0001-8043-8413]{David P. Bennett}
\affiliation{Department of Astronomy, University of Maryland, College Park, MD 20742, USA}
\affiliation{Code 667, NASA Goddard Space Flight Center, Greenbelt, MD 20771, USA}
\email{bennett.moa@gmail.com}

\author{Aparna Bhattacharya}
\affiliation{Department of Astronomy, University of Maryland, College Park, MD 20742, USA}
\affiliation{Code 667, NASA Goddard Space Flight Center, Greenbelt, MD 20771, USA}
\email{abhatta5@umd.edu}

\author[0000-0002-8131-8891]{Ian A. Bond}
\affiliation{School of Mathematical and Computational Sciences, Massey University, Auckland 0632, New Zealand}
\email{I.A.Bond@massey.ac.nz}

\author[0009-0007-3944-7298]{Jon Hulberg}
\affiliation{Code 667, NASA Goddard Space Flight Center, Greenbelt, MD 20771, USA}
\affiliation{Department of Physics, Catholic University of America, Washington, DC 20064, USA}
\affiliation{Center for Research and Exploration in Space Science and Technology, NASA/GSFC, Greenbelt, MD 20771}
\email{brashearj@cua.edu}

\author[0000-0003-2267-1246]{Stela {Ishitani Silva}}
\affiliation{Code 667, NASA Goddard Space Flight Center, Greenbelt, MD 20771, USA}
\affiliation{IPAC, Caltech, 1200 E. California Blvd., Pasadena, CA 91125, USA}
\email{stela@ipac.caltech.edu}

\author[0000-0001-7016-1692]{Przemek Mróz}
\affiliation{Astronomical Observatory, University of Warsaw, Al. Ujazdowskie 4, 00-478 Warszawa, Poland}
\email{pmroz@astrouw.edu.pl}

\author[0000-0002-9881-4760]{Aikaterini Vandorou}
\affiliation{Department of Astronomy, University of Maryland, College Park, MD 20742, USA}
\affiliation{Code 667, NASA Goddard Space Flight Center, Greenbelt, MD 20771, USA}
\email{katievan@umd.edu}

\correspondingauthor{T. Dex Bhadra}

\begin{abstract}

\small \noindent We present the Microlensing Object high-Resolution Imaging Analysis pipeline, or \texttt{MORIA}. This is an automated procedure to reduce high-resolution \textit{HST} images of microlensing targets, build empirical point-spread function models from the data, and perform simultaneous multi-star PSF fitting to blended sources, lenses, and neighbor stars. We have developed and tested this pipeline using HST observations of the microlensing event KMT-2019-BLG-0253, which shows clear evidence of three highly-blended stars in HST. Under our assumed source and lens identification, we determine a host mass of $M_{host} = 0.65 \pm 0.04M_{\odot}$. We have reduced the number of possible solutions for this target by a factor of two, with the remaining solution subject to the well known close-wide degeneracy. We determine a planet mass of $m_{p} = 7.18 \pm 0.40 M_{\oplus}$ (close) or $m_{p} = 9.48 \pm 1.13 M_{\oplus}$ (wide), and distance to the lens system of $D_L= 2.64 \pm 0.22$ kpc. This target lies within the upcoming Roman Telescope Galactic Bulge Time Domain Survey (GBTDS) field. Future analysis of the Roman data for this event can confirm or reject the initial source and lens identifications presented in this work. Finally, this work demonstrates the importance of using an automated high-resolution imaging tool to inform light curve modeling for microlensing planets found during the GBTDS.
\end{abstract}


\section{Introduction} \label{sec:intro}

Gravitational microlensing is unique among the exoplanet detection methods because of its ability to detect cold exoplanets \citep{mao:1991a,gould:1992a}, and particularly low-mass planets beyond the snow line \citep{bennett:1996a}. One drawback of this method is that for most microlensing lightcurves, only the mass-ratio ($q$) of the lens system is measured, which leaves some physical parameters of the system significantly unconstrained. This can result in large estimated uncertainties, particularly in the inferred stellar host and companion masses due to uncertain priors used in the standard Bayesian modeling approach.  One can mitigate this limitation either by measuring multiple ``higher-order effects" in the light curve data, typically difficult in ground-based observations, or by resolving the source and lens independently with high angular resolution imaging (i.e. \textit{Hubble Space Telescope} (HST), Keck AO, Subaru AO) several years after peak magnification \citep{kozlowski:2007a,bennett:2006a,bennett:2007a}. This high angular resolution imaging allows one to further constrain the lens-source separation, relative proper motion between the targets, and lens flux which can then be used with mass-luminosity relations \citep{henry:1993a,henry:1999a,delfosse:2000a} to infer a direct mass for the host and its companion. \\
\indent At present, the microlensing technique has produced over 270 published exoplanets at distances up to the Galactic Bulge\footnote{\url{https://exoplanetarchive.ipac.caltech.edu/}}. Although this number is less than transit or radial velocity discoveries, occurrence rate studies have been performed on the microlensing exoplanet population thus far \citep{mroz:2023}. As for most transient phenomena, one limitation of this method for fully characterizing microlensing systems is the cadence at which the photometric data is obtained by the dedicated ground-based surveys. Two of the longest-operating surveys are the Optical Gravitational Lensing Experiment (OGLE \citep{udalski:2015a}) with one $1.3$ m telescope based in Chile, and Microlensing Observations in Astrophysics (MOA \citep{bond:2001a}) with one $1.8$ m telescope based in New Zealand. In 2016 the Korea Microlensing Telescope Network (KMTNet; \cite{kim:2016a}) began monitoring of the Galactic bulge with a network of three ground-based telescopes located in Australia, South Africa, and Chile. This network has enabled high-cadence, nearly continuous monitoring of the stars toward the bulge, which has resulted in a measurable increase in the number of published microlensing planets from a {$\sim$}few per year last decade to over 10 per year currently \citep{mroz:2023a}. \\
\indent NASA's Nancy Grace Roman Space Telescope (\textit{Roman}) is currently scheduled to launch in Fall 2026 and will conduct the \textit{Roman} Galactic Bulge Time Domain Survey (\textit{GBTDS}) {\citep{Gaudi:2022a}}. As part of this bulge survey, the Roman Galactic Exoplanet Survey (\textit{RGES}) will be the first dedicated space-based gravitational microlensing survey and is expected to detect over 1400 bound exoplanets \citep{penny:2019a, zohrabi:inprep}, hundreds of free-floating planets \citep{johnson:2020a, sumi:2023a}, compact objects \citep{lam:2020a, sajadian:2023a} and more during its five-year survey. This mission will complement previous large statistical studies of transiting planets from missions like \textit{Kepler}/\textit{TESS} and radial velocity (RV) planets from many ground-based RV surveys. One significant benefit of conducting a microlensing survey from space is to achieve high-precision, high-cadence photometry of lightcurves, which can aid in the measurement of higher-order effects mentioned earlier. Additionally, \textit{Roman} itself can act as a high angular resolution followup facility to attempt direct lens flux detections for microlensing events it discovers \citep{bhattacharya:2018a}. Recently \cite{terry:2026a} performed forward modeling of 3,000 simulated \textit{Roman} planetary microlensing events using \texttt{pyLIMASS} \citep{bachelet:2024a} to predict the fraction of events for which lens masses and distances can be precisely measured (within 20\% uncertainty). They found ${\sim}$40\% of total events meet the 20\% uncertainty threshold in both mass and distance estimates. \\
\indent While \texttt{pyLIMASS} is a powerful tool for estimating lens system physical properties in microlensing events via Gaussian mixture modeling, there does not exist a publicly-available, automated  end-to-end pipeline to reduce high angular resolution imaging data and perform lens flux analyses (i.e. multiple-component point spread function fitting) together. In this paper, we present the Microlensing Object high-Resolution Imaging Analysis (\texttt{MORIA}) pipeline. First, in Sec. \ref{sec:event}, we present the original observations of the microlensing event KMT-2019-BLG-0253. Sec. \ref{sec:followup} describes archival imaging that serendipitously observed the target with \textit{HST} in 2005, as well as our followup imaging with \textit{HST} obtained in 2025. Our new imaging reveals the presence of three stars at the approximate location of the target, which leads us to consider multiple interpretations for the source and lens systems (i.e. binarity). We also perform updated modeling of the lightcurve data in this section to determine the nature of the three stars. Once we identify the lens and the source in Sec.~\ref{sec:followup}, we report our estimates of the lens-source separation, flux ratio, and the relative proper motion in Sec.~\ref{sec:prop-motion}. In Sec.~\ref{sec:lens-properties} we present lens system properties with new constraints from high-resolution imaging. We introduce our automated pipeline \MORIA for high-resolution image analysis and its application to KMT-2019-BLG-0253 in Sec.~\ref{sec:automated_pipeline}, as well as its broader application for future \textit{Roman} events in Sec.~\ref{sec:roman_moria}. Finally, we discuss the overall results and conclude the paper in Sec. \ref{sec:conclusion}.


\section{Microlensing Event KMT-2019-BLG-0253} \label{sec:event}

KMT-2019-BLG-0253 (hereafter KB190253), located at RA $=$ 17:51:31.82, DEC $=$ -29:33:55.7 and Galactic coordinates ($\ell,b=(0.126, -1.429)$) was alerted by all three major ground-based microlensing surveys; OGLE, MOA, and KMTNet. The event was originally published in \cite{hwang:2022a} which included six total planetary events with mass ratios of $q < 2\times10^{-4}$. We note the event also has the identifiers OGLE-2019-BLG-0410 and MOA-2019-BLG-127.\\
\indent \citet{hwang:2022a} presented four non-static solutions for KB190253. Our work in this study, presented below, reduces these degeneracies by a factor of two. Further, the event was reported to have significant blended light. It was not definitive whether the blend was due to the lens itself or a companion to the lens, source, or an ambient field star. In our high-resolution followup analysis we confirm three highly-blended stars at the approximate position of the target in HST. While our analysis in this study prefers `Star 2' as the source and `Star 3' as the lens, an additional epoch of high-resolution imaging in the future can confirm or reject our identifications and will be critical for determining the true nature of the detected objects (see Sections \ref{sec:multi-star} and \ref{sec:source-lens}). \\
\indent \cite{hwang:2022a} use a Bayesian analysis to conclude that KB190253 has a super-Earth mass planet (${\sim}9\, M_{\oplus}$) orbiting a $0.7\, M_{\odot}$ host at a distance of 5 kpc from Earth.


\section{High-resolution precursor \& followup imaging with HST} \label{sec:followup}

\subsection{2005 $HST$ Observations}\label{hst-analysis-2005}
\begin{figure} [!htb]
\centering
\includegraphics[width=0.9\linewidth]{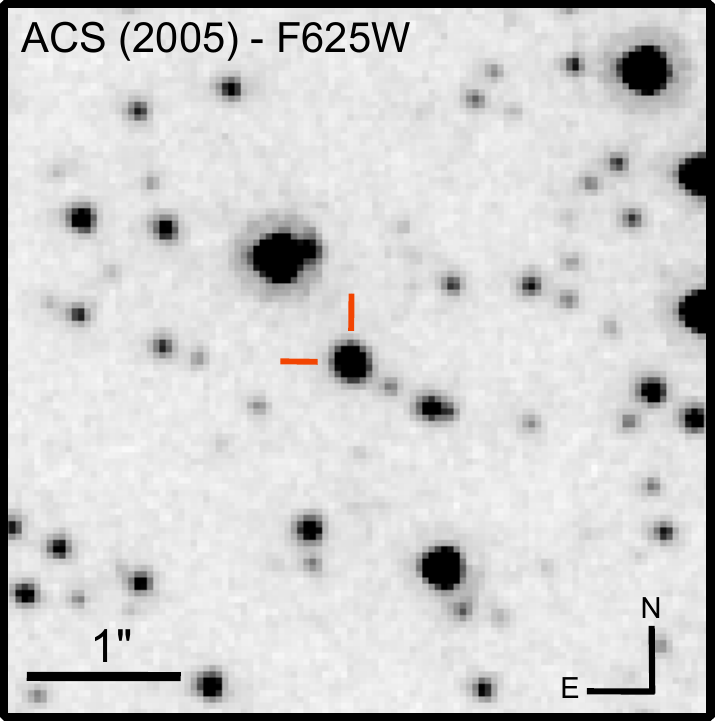}

\centering
\caption{A zoom-in on the target from 2005 ACS imaging in the $F625W$ filter. The target location was confirmed through astrometric analysis of the magnified source centroid in OGLE images. The yellow data point shows the source location transformed from OGLE coordinates to the HST image, with error bars corresponding to the statistical uncertainty of the centroid as propagated onto the HST frame and the rms of the transformation itself.} \label{fig:2005_target}
\end{figure}

The \textit{HST} program GO-10353 \citep{grindlay:2005prop} conducted a deep survey of low-luminosity accretion-powered sources including cataclysmic variables (CVs), low-mass x-ray binaries (LMXBs), and wind-fed pulsars. These images, taken with the Wide-Field Camera (WFC) module of the Advanced Camera for Surveys (ACS), serendipitously captured KB190253 in the field of view (FoV) with several passbands. \\
\indent This program was conducted in 2005, approximately 16.6 years before the microlensing event occurred, and the pixel scale of WFC is 50 mas/pixel. Figure \ref{fig:2005_target} shows a zoom on the drizzled $ACS$ image in 2005, with the target(s) at center frame indicated by a red cross-hair. There is an apparent elongation in the stellar profile, which indicates a blend of more than one point source (i.e. the source and lens). The field has relatively high crowding, with approximately eight stars brighter than $F625W < 23$ per sq. arcsecond along this sight-line.\\
\indent Given the nature of this archival dataset, the observing strategy (e.g. dithering scheme, exposure times, etc.) was not optimized to perform the high precision astrometry that is typical of our previous high-resolution followup analysis for microlensing events \citep{bennett:2015a, bhattacharya:2018a, terry:2024a}. At present, \texttt{MORIA} is built to function on WFC3/UVIS data only, thus we do not perform a multiple-star PSF fitting procedure on this 2005 ACS/WFC dataset. However future development of \texttt{MORIA} can include support for this ACS/WFC instrument and more (see Sec. \ref{sec:future_moria}).

\subsection{2025 HST Observations \& Multi-star PSF Fitting}\label{sec:multi-star}

\begin{figure*}[!htb]
\includegraphics[width=0.8\linewidth]{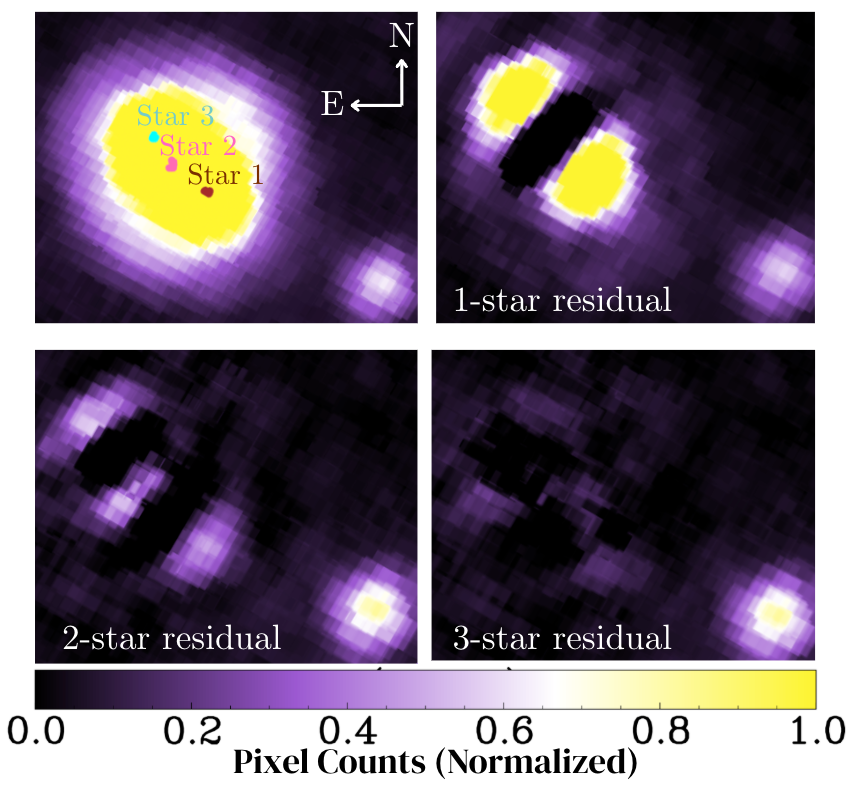}
\centering
\caption{\footnotesize \emph{Top left:} Zoomed \textit{HST} $I-$band image of the source, lens, and blend stars, along with the MCMC chains centering on their positions, found using \MORIA. The 1D separation between Star 2 (assumed source) and Star 3 (assumed lens) in this epoch is 48.13 mas. \emph{Top right:} The residual image from a single-star PSF fit. A clear signal is seen due to the blended stellar profiles. \emph{Bottom left:} The residual image from a two-star PSF fit. A significant improvement is seen, compared to the single-star PSF fit. However, there remains a coherent signal that cannot be solely attributed to noise. \emph{Bottom right:} The residual image from a triple-star PSF fit, which shows a significantly improved subtraction. The color bar represents the normalized pixel intensity (or counts) in the residual image panels. \label{fig:4panel_hst}}
\end{figure*}

Approximately 19.5 years after the serendipitous observations of \cite{grindlay:2005prop}, the program GO-17834 \citep{terry:2024prop} conducted dedicated observations of KB190253 with the WFC3/UVIS camera on 26 April 2025. UVIS has a pixel scale slightly better than ACS/WFC, at 40 mas/pixel, and employs a similar suite of filters. For the 2025 observations, a total of 26 successful \textit{HST} exposures were taken, 10 in the $F814W$ (wide I$-$band) filter and 16 in the $F606W$ (wide V$-$band). Six of the $F814W$ exposures failed due to a guiding fault in reduced-gyro mode (RGM) onboard \textit{HST}.\\
\indent Our data reduction and PSF fitting analysis was performed using an automated version of the procedure originally developed in \citet{bennett:2015a}; we call this automated analysis \texttt{MORIA}. Further detail on our development of \texttt{MORIA}, including each major component of the routine, is given in Section \ref{sec:automated_pipeline}. We present the results of our multiple-star PSF fitting for KB190253 in Figure~\ref{fig:4panel_hst}. The top-left panel shows the target and the surrounding HST stars from the $I-$band image taken in 2025. The top-right panel shows the residual image after fitting a single-star PSF model. The clear dipole structure seen in the target demands a multiple-star PSF fit. In the bottom left panel, the two-star residuals are an improvement over a single-star fit, however the presence of structural noise remains which motivates the need for a simultaneous three-star fit. Lastly, we present the three-star residual image in the bottom-right panel of Figure \ref{fig:4panel_hst}. This three-PSF model fit gives a $\chi ^2$ improvement over the single-star and double-star PSF fits of $\chi ^2_{\textrm{1star}} - \chi ^2_{\textrm{3star}} = 346.17$ and $\chi ^2_{\textrm{2star}} - \chi ^2_{\textrm{3star}} = 41.23$, respectively.\\
\indent At this stage in the analysis there is no definitive determination of which of the three blended stars is the source or lens. However Star 2's best-fit magnitude in $I-$band matches most closely to the fitted source magnitude from the original lightcurve modeling of \cite{hwang:2022a}. The original study fit this microlensing event with a 2L1S model, and the authors indicate evidence for a blend object within ${\sim}50$ mas of the source at the time of the microlensing event. Our \textit{HST} observations confirm the presence of three blended objects, however definitively confirming the nature of the third object will require at least one additional epoch of high-resolution imaging to measure the per-epoch motion of all three objects. Our reduction and fitting code places the star coordinates from both filters into the same reference system, so all stars have positions that are consistent between both passbands.\\
\indent The HST data were calibrated to the OGLE-III catalog \citet{szymanski:2011a} using nine relatively bright isolated OGLE-III stars that were matched to HST stars. For the 2025 \textit{HST} epoch, the calibrations yield $I_{\textrm{star1}} = 20.296 \pm 0.060$, $I_{\textrm{star2}} = 19.685 \pm 0.049$, $I_{\textrm{star3}} = 19.941\pm 0.054$, $V_{\textrm{star1}} = 23.433\pm 0.091$, $V_{\textrm{star2}} =22.712 \pm 0.072$, $V_{\textrm{star3}} = 22.834 \pm 0.075$. These calibrated magnitudes for each star are also presented in Table~\ref{tab:dual-phot}. Lastly, the high-precision magnitudes enabled by our \textit{HST} observations allow us to place strong constraints when re-modeling the microlensing lightcurve photometry.

\begin{deluxetable}{lll}
\deluxetablecaption{\textit{HST} 2025 Multi-star PSF photometry\label{tab:dual-phot}}
\tablecolumns{3}
\setlength{\tabcolsep}{12pt}
\tablewidth{0pt}
\tablehead{
\colhead{Object} &
\colhead{$V$ Mag} & \colhead{$I$ Mag}
}
\startdata
Star 1 & $23.433 \pm 0.091$ & $20.296 \pm 0.060$\\
Star 2 & $22.712 \pm 0.072$ & $19.685 \pm 0.049$\\
Star 3 & $22.834 \pm 0.075$ & $19.941 \pm 0.054$\\
\enddata
\tablenotetext{}{\footnotesize{\textbf{Note}. $V$ and $I$ magnitudes are calibrated to the OGLE-III system as described in Sec.~\ref{sec:followup}.}}
\end{deluxetable}

\subsection{Lightcurve Fitting \label{sec:light-curve}}

\cite{bhattacharya:2018a, bennett:2020a, terry:2021a, rektsini:2024a} and other recent studies have shown it can be particularly useful to apply constraints from the high-resolution followup observations to the microlensing lightcurve models (deemed ``image-constrained modeling"). This can help prevent the lightcurve modeling from exploring areas in the parameter space that are excluded by the high-resolution followup observations. We refer the reader to \cite{bennett:2023a} for a full description of the methodology for applying these constraints to the modeling.\\
\indent We use a modified version of the lightcurve modeling code \texttt{eesunhong} \citep{bennett:1996a, bennett:2010a} to incorporate constraints on the brightness and separation of the lens and source stars from the high-resolution imaging via \textit{HST}. Ideally, we want to use a mass-distance relation coupled with empirical mass-luminosity relations to infer the mass and distance of the host star. In order to do this, we need to know the distance to the source star, $D_S$. Thus we are required to include the source distance as a fitting parameter in the re-modeling of the lightcurve with imaging constraints. We include a weighting from the \cite{koshimoto:2021a} Galactic model as a prior for $D_S$, and we also use the same Galactic model to obtain a prior on the lens distance for a given value of $D_S$. This prior is not used directly in the lightcurve modeling, but instead is used to weight the entries in a sum of Markov chain values.

\begin{deluxetable*}{llccc|c}
\tablecaption{Measured Lens--Source Separations from \textit{HST} for Star 1, Star 2, and Star 3\label{tab:combined_sep}.}
\tablecolumns{6}
\setlength{\tabcolsep}{6pt}
\tablewidth{\textwidth}
\tablehead{
\colhead{Pair} &
\colhead{Filter} &
\multicolumn{3}{c}{Separation (mas)} &
\colhead{$\Delta\chi^2_{\rm L\leftrightarrow S}$}
\\ 
\cline{3-5}
 & & \colhead{North} & \colhead{East} & \colhead{Total} & 
}
\startdata
Star 1 (Lens) \& Star 2 (Source) & (\textit{HST I}) & $-31.45 \pm 0.53$ & $-47.82 \pm 0.56$ & $57.23 \pm 0.55$ \\
 & (\textit{HST V}) & $-30.58 \pm 0.25$ & $-46.58 \pm 0.24$ & $55.73 \pm 0.24$ \\
\cline{2-5}
 & & $\mu_{\rm rel,H,N}$ (mas/yr) & $\mu_{\rm rel,H,E}$ (mas/yr) & $\boldsymbol{\mu}_{\rm rel,H}$ (mas/yr) & 145\\
\cline{2-5}
 & weighted mean & $-5.53 \pm 0.04$ & $-7.79 \pm 0.04$ & $-9.33 \pm 0.04$ \\
\hline\hline
Star 1 (Lens) \& Star 3 (Source) & (\textit{HST I}) & $-49.94 \pm 0.20$ & $-92.24 \pm 0.20$ & $104.88 \pm 0.20$ \\
 & (\textit{HST V}) & $-47.17 \pm 0.13$ & $-93.26 \pm 0.14$ & $104.55 \pm 0.14$ \\
\cline{2-5}
& & $\mu_{\rm rel,H,N}$ (mas/yr) & $\mu_{\rm rel,H,E}$ (mas/yr) & $\boldsymbol{\mu}_{\rm rel,H}$ (mas/yr) & 760 \\
\cline{2-5}
 & weighted mean & $-7.96 \pm 0.02$ & $-15.51 \pm 0.02$ & $-17.44 \pm 0.02$ \\
\hline\hline
\textbf{Star 2 (Source) \& Star 3 (Lens)}& (\textit{HST I}) & $20.78 \pm 0.57$ & $43.41 \pm 0.70$ & $48.13 \pm 0.65$ \\
 & (\textit{HST V}) & $18.99 \pm 0.13$ & $45.75 \pm 0.14$ & $48.31 \pm 0.14$ \\
\cline{2-5}
 & & $\mu_{\rm rel,H,N}$ (mas/yr) & $\mu_{\rm rel,H,E}$ (mas/yr) & $\boldsymbol{\mu}_{\rm rel,H}$ (mas/yr) & 115 \\
\cline{2-5}
 & weighted mean & $3.45 \pm 0.09$ & $7.21 \pm 0.12$ & $7.98 \pm 0.21$ \\
\enddata
\tablenotetext{}{\footnotesize{\textbf{Notes.} The table presents our three trial combinations of lens and source. The highlighted row indicates the pair that yields the best $\chi^2$ via image-constrained lightcurve fitting (see Sec. \ref{sec:light-curve}). The last column ($\Delta\chi^2_{\rm L\leftrightarrow S}$) represents the $\chi^2$ improvement of the given Lens/Source `Pair' over its inverse scenario, as given by the image-constrained light curve models. A $\Delta\chi^2_{\rm L\leftrightarrow S} > 0$ indicates that the Lens-Source `Pair' defined in column 1 is favored over its inverse (Source-Lens) Pair.}}
\end{deluxetable*}

\indent In Figure~\ref{fig:4panel_hst}, the MCMC chains for the three possible stars are displayed in the top-left panel. Based on the three possible pairs of stars, there are six different brightness, separation and source-lens proper motion ($\boldsymbol{\mu}_{\rm rel,H}$) constraints for all six possible pairs of star combinations. Using the $\chi^2$ results from \texttt{eesunhong}, we eliminated all possibilities where the brighter star was chosen as the trial lens during the lightcurve fitting process. The separation and $\boldsymbol{\mu}_{\rm rel,H}$ for the remaining three possibilities, where the fainter star was chosen as the trial lens during the lightcurve fitting process, are summarized in Table~\ref{tab:combined_sep}. The final column in Table~\ref{tab:combined_sep} reports $\Delta \chi_{imp}^2$, which is the $\Delta \chi^2$ improvement of the reported source-lens assignment in the first column relative to the inverse assignment of the same stellar pair. For example, Star 2 as source and Star 3 as lens is favored by $\Delta\chi^2 = 115$ over Star 2 as lens and Star 3 as source. The corresponding $\Delta\chi_{imp}^2$ for Star 1 as source and Star 2 as lens is $145$; the $\Delta\chi_{imp}^2$ for Star 1 as source and Star 3 as lens is $760$. The measured lens-source separation point from the source (brighter) to the lens (fainter), as indicated by the sign convention on $\boldsymbol{\mu}_{\rm rel}$ in Table~\ref{tab:combined_sep}. However, note that Table~\ref{tab:combined_sep} does not actually determine the true lens and source in this study; it merely indicates which star was assumed as the lens/source for each pair to feed into the image-constrained modeling code \texttt{eesunhong}. \\
\indent The results from \texttt{eesunhong} are as follows. The two-lens, one-source (2L1S)  model for Star 2 as the source and Star 3 as the Lens returned the best fit $\chi^2$ with an improvement of $\Delta \chi^2 \approx 3196 $ over the Star 1 (lens) and Star 3 (source) run; and an improvement of $\Delta \chi^2 \approx 608 $ over the Star 1 (lens) and Star 2 (source) run. Furthermore, the 2L1S model for the Star 2-Star 3 combination had an improvement of $\Delta \chi^2 \approx 615$ from the best-fit image-constrained two-lens, two-source (2L2S) models. It also had an improvement of $\Delta \chi^2 \approx 1951$ over the one-lens, one-source (1L1S) model. Additionally, the best-fit 2L1S model had a positive $u_0$, and negative $\alpha$; it had a $\Delta \chi^2 \approx 158$ improvement over the elliptically degenerate case with negative $u_0$ and positive $\alpha$. The total $\chi^2$ values reported by \texttt{eesunhong} are computed as the sum of contributions from the lightcurve fit and from image-based constraints, including those derived from the source, lens, and source+lens $I-$ and $V-$ band magnitudes, as well as the $\boldsymbol{\mu}_{\rm rel, H}$ constraint. 

\begin{figure*}
\includegraphics[width= \textwidth]{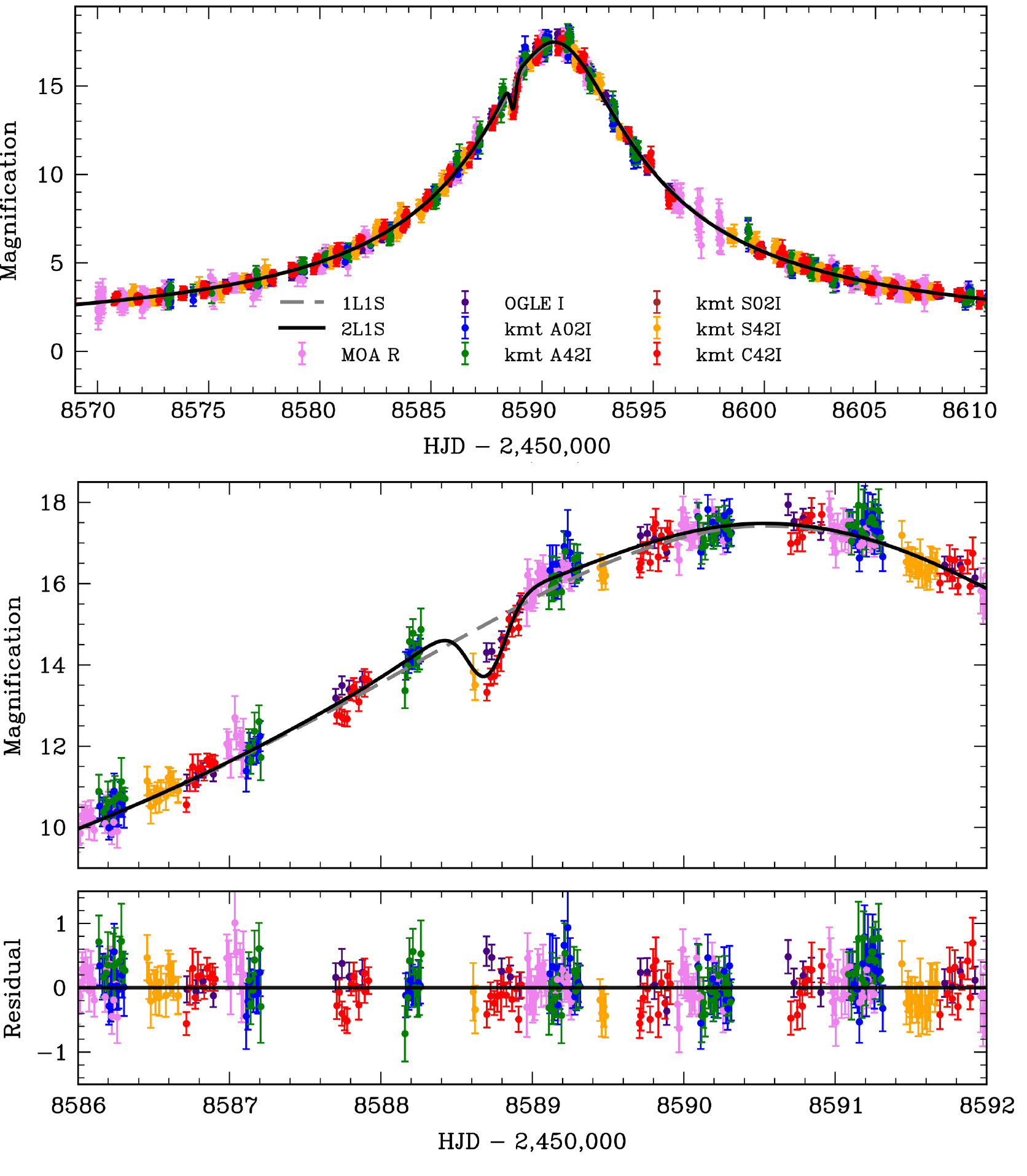}
\centering
\caption{\footnotesize Best-fit lightcurve with constraints from the high--resolution followup data as described in Sec.~\ref{sec:followup}. The 2L1S model shown here is the close solution given in Table~\ref{tab:lcpar_combined}. Note that it is identical to the wide solution. The top panel shows a wide view of the overall magnification event. The middle panel shows a zoom-in around the planetary anomaly. The bottom panel shows the residuals of the 2L1S fit. \label{fig:lc}}
\end{figure*}

\indent Figure~\ref{fig:lc} shows the observed lightcurve with OGLE, MOA, and KMTNet data, as well as the best-fit planetary model (2L1S) from our re-analysis of the lightcurve modeling with high-resolution imaging constraints applied. The best 1L1S model is also shown for comparison.

\subsection{Identifying the Source and Lens\label{sec:source-lens}}

\begin{figure*}
\centering
\includegraphics[width=0.75\linewidth]{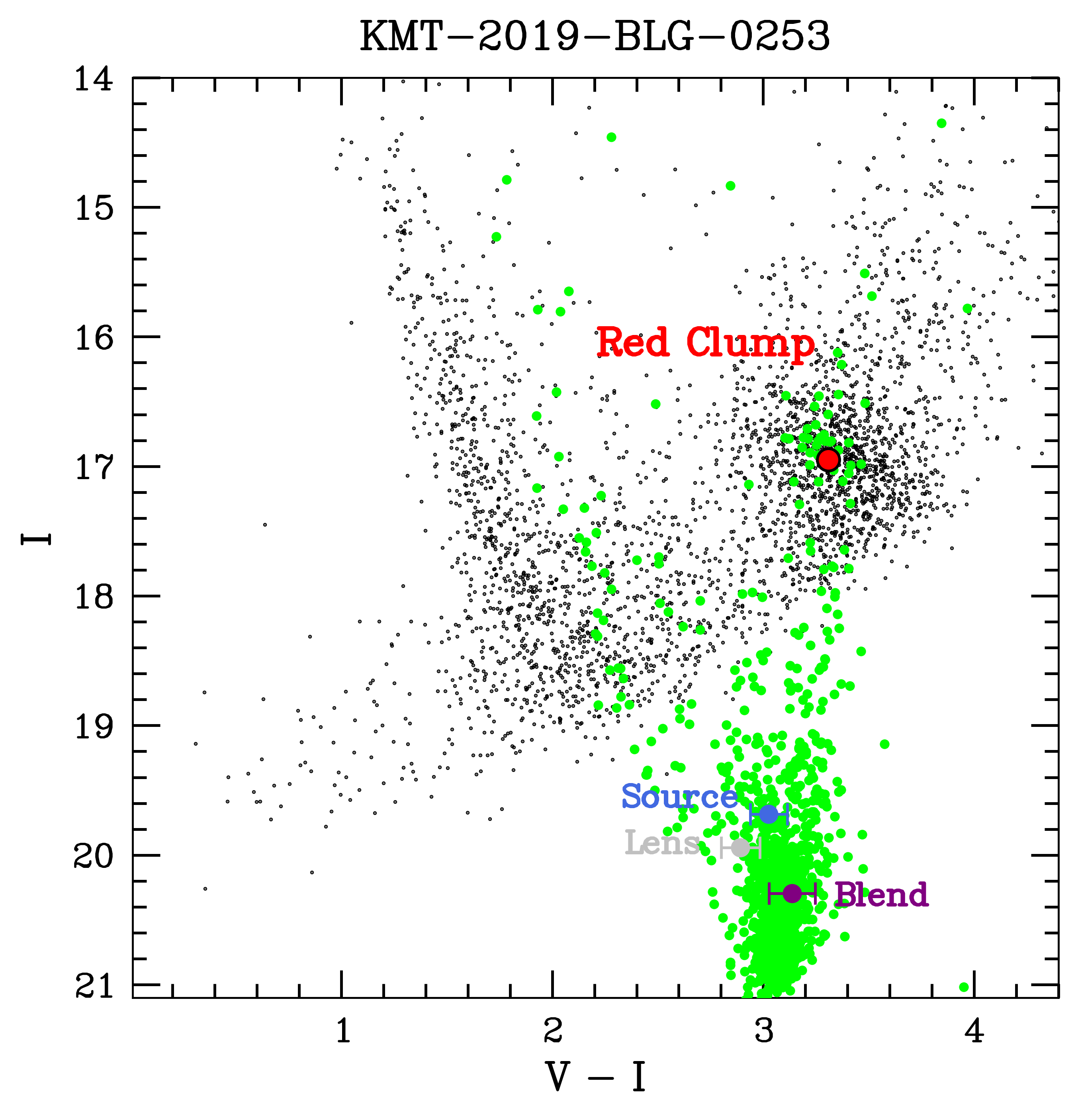}

\centering
\caption{The observed color–magnitude diagram (CMD) for the KB190253
field. The OGLE-III stars within 90" of KB190253 are shown in black, with the HST CMD of all detected sources from the 2025 epoch shown in green. The
red point indicates the location of the red clump centroid. The blue, grey, and purple points show the colors and magnitudes of the three stars (source, lens, blend) measured in \textit{HST}. \label{fig:CMD}}
\end{figure*}

Based on our lightcurve modeling (with and without high-res imaging constraints), we find strong evidence that Star 2 is the source and Star 3 is the lens (host star). Although our 2L2S is significantly disfavored from the image-constrained lightcurve modeling, at present we cannot definitively rule out Star 1 as a companion to the source (or lens). This will require at least one additional epoch of high-resolution imaging to measure the proper motion of each object to determine if Star 1's motion is coincident with either of the other two stars. \\
\begin{figure}
\includegraphics[width= 0.45\textwidth]{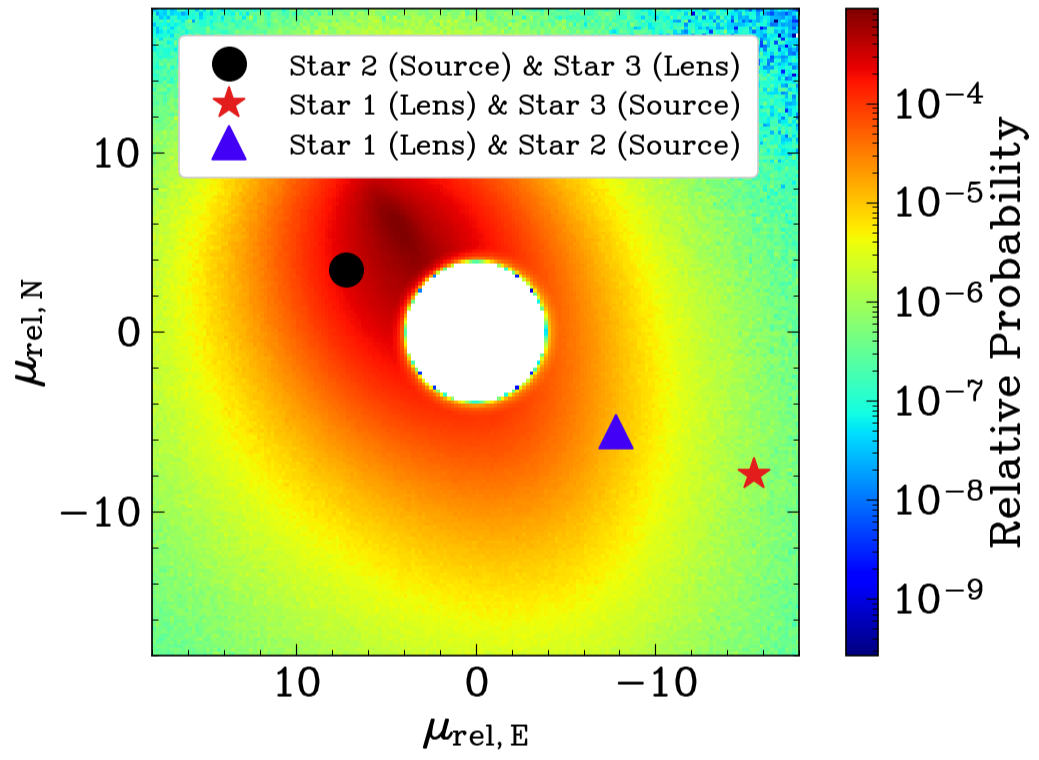} 
\centering
\caption{\label{fig:mulens}The probability distribution for the north and east components of
lens–source relative proper motion ($\boldsymbol{\mu}_{\rm rel}$) using the Galactic model from \citet{koshimoto:2021a} and \texttt{genulens} \citep{koshimoto:code} The three possible combinations of source and lenses between star 1, 2 and 3 are plotted in black red and purple. The values are given by the relative motion of the two stars detected in HST in Table~\ref{tab:combined_sep}.}
\end{figure}
\indent As an additional test, we calculate the 2D prior probability distribution of the lens-source relative proper motion ($\boldsymbol{\mu}_{\rm rel}$) using the \citet{koshimoto:2021a} Galactic model to determine which combination of stars is the preferred lens and source pair. Figure~\ref{fig:mulens} shows this proper-motion distribution for KB190253, with three locations for the possible $\boldsymbol{\mu}_{\rm rel}$ values depending on which star is assumed as the source and which star is the assumed lens in Table~\ref{tab:combined_sep}. The results show that the Galactic model clearly prefers ``Star 2" as the source and ``Star 3" as the lens. The relative probability for this pair is $>$$10\times$ more likely than the Star 1-Star 2 pair and $>$$100\times$ more likely than the Star 1-Star 3 pair. With this Galactic model evidence and the lightcurve modeling results from earlier, we proceed under the identification of ``Star 2" as the source, ``Star 3" as the lens, and ``Star 1" as the blend (at present). These identifications are reflected in the calibrated color-magnitude diagram (CMD) shown in Figure~\ref{fig:CMD}.

\subsection{Extinction Toward the Source and Lens Star\label{sec:extinction}}
The OGLE extinction calculator is a standard way to estimate the extinction for a Galactic bulge field, and it has been commonly used for many years. The calculator is derived from the reddening and extinction study of \citet{nataf:2013a}. For the KB190253 sight line, the estimated $I-$band extinction to the source is $A_{I,s}=2.75$, with a reddening of $E(V-I)=2.30$. Combining equation 21 of \cite{nataf:2013a} with the assumption that the dust scale height is $165$ pc, we derive an extinction to the lens system of $A_{I,l}\sim0.54$, $A_{I,s}{\sim} 1.49$. This is consistent with a relatively nearby lens system ($< 4$ kpc) and the moderately high reddening toward this sight line.

\section{Lens-Source Relative Proper Motion} \label{sec:prop-motion}
The 2025 \textit{HST} followup observations were taken 6.02 years after peak magnification in 2019. The motion of the lens and source on the sky frame is the primary cause for their apparent separation. However the third star accompanying the blended stellar profile adds an additional level of complexity, as explained in Sec.~\ref{sec:source-lens}.\\
\indent Following from our identification of the lens and source stars, we measure a mean lens--source relative proper motion of $\boldsymbol{\mu}_{\rm rel,H} = (\mu_{\rm rel,H,E}, \mu_{\rm rel,H,N}) = (7.21 \pm 0.12,\, 3.45 \pm 0.09)\ \mathrm{mas\ yr^{-1}}$, where the ``H'' subscript indicates that these measurements were made in the heliocentric reference frame, and the ``E'' and ``N'' subscripts represent the east and north on-sky directions, respectively. Our lightcurve modeling is performed in the geocentric reference frame that moves with the Earth at the time of the event peak. Thus, we must convert between the geocentric and heliocentric frames by using the relation given by \citet{dong:2009a}.

\begin{equation}
\boldsymbol{\mu}_{\rm rel,H} = \boldsymbol{\mu}_{\rm rel,G} + \frac{\pi_{\rm rel}}{\mathrm{au}} \, \boldsymbol{v}_{\oplus},
\end{equation}

where $\boldsymbol{v}_{\oplus}$ is Earth’s projected velocity relative to the Sun at the time of peak magnification. For KB190253, this value is $\boldsymbol{v}_{\oplus, E,N} = (2.589,\, 14.165)\ \mathrm{km\ s^{-1}} = (0.546,\, 2.985)\ \mathrm{au\ yr^{-1}}$ at $\mathrm{HJD}' = 8590.60$, where $\mathrm{HJD}' = \mathrm{HJD} - 2450000$. With this information and the relative parallax relation $\pi_{\rm rel} \equiv \mathrm{au}\left(1/D_L - 1/D_S\right)$, we can express Equation~(2) in a more convenient form:

\begin{equation}
\label{eqn:two}
\boldsymbol{\mu}_{\rm rel,G} = \boldsymbol{\mu}_{\rm rel,H}
- (0.546,\, 2.985)\left(\frac{1}{D_L} - \frac{1}{D_S}\right)
\ \mathrm{mas\ yr^{-1}},
\end{equation}

where $D_L$ and $D_S$ are the lens and source distances, respectively, given in kiloparsecs.\\
\indent As stated in Sec. \ref{hst-analysis-2005}, we are currently unable to perform multiple-star PSF fitting of the 2005 \textit{HST/ACS} data. We perform a brief cross-validation of the relative proper motion values we estimate based on differing assumptions of lens and source identification (Table \ref{tab:dual-phot}). The largest estimate of relative motion from Table \ref{tab:dual-phot}, $\boldsymbol{\mu}_{\rm rel,H} = (-17.44 \pm 0.02)\ \mathrm{mas\ yr^{-1}}$, would have a low likelihood based on lens and source star kinematics via Galactic models \citep{koshimoto:2021a}. Propagating this relative proper motion back to the 2005 epoch would result in a lens-source separation of nearly ${\sim 6}$ pixels in \textit{HST/ACS}. Visual inspection of the target in the 2005 data (e.g. Figure \ref{fig:2005_target}) shows slight elongation of the stellar profile, but is not significant enough to indicate a separation as large as a ${\sim 17}$ mas/yr proper motion would imply. Another complication is that we know there are three point sources blended together in the 2005 \textit{HST/ACS} image based on our analysis of the 2025 \textit{HST/WFC3} images.\\
\indent We have directly measured $\boldsymbol{\mu}_{\rm rel,H}$ from the \textit{HST} data, so this gives us the relative proper motion in the geocentric frame of $\mu_{\rm rel,G} = 7.58 \pm 0.19\ \mathrm{mas\ yr^{-1}}$. As a reminder, the lens and source distances we use in Equation~\ref{eqn:two} are inferred by the best-fit lightcurve results, which include constraints from the high-resolution imaging.

\section{Lens System Properties} \label{sec:lens-properties}

Table~\ref{tab:lcpar_combined} summarizes the parameters of the two best-fit solutions, as well as the MCMC averages of models consistent with the data. The inclusion of the HST constraints substantially reduces the uncertainties of several model parameters compared to the light-curve-only analysis of \citet{hwang:2022a}. The HST imaging directly constrains the source and lens fluxes and measures the heliocentric lens-source relative proper motion, $\boldsymbol{\mu}_{\rm rel,H}$.The lightcurve modeling is performed in the geocentric frame, so the measured heliocentric relative proper motion must be transformed to $\boldsymbol{\mu}_{\rm rel,G}$ (Eq. \ref{eqn:two}). The HST source photometry constrains $\theta_*$ through the adopted extinction and color-surface-brightness relation, rather than providing a direct measurement of $\theta_*$. Together, these imaging constraints restrict the allowed values of $t_*=\theta_*/\boldsymbol{\mu}_{\rm rel,G}$ and result in a substantially smaller uncertainty on $t_*$ than in the lightcurve-only analysis of \citet{hwang:2022a}. The proper-motion constraint also restricts the individual microlensing parallax components because the parallax vector is parallel to $\boldsymbol{\mu}_{\rm rel,G}$. From the four solutions first reported in \citet{hwang:2022a}, our image-constrained modeling has reduced the number of solutions by 2. We have ruled out the two $u_0 < 0$ solutions from \citet{hwang:2022a}; the remaining solution is subject to the close/wide degeneracy.

\begin{deluxetable*}{lcccccc}[t]
\deluxetablecaption{Best Fit KMT-2019-BLG-0253 Model Parameters\label{tab:lcpar_combined}}
\tablecolumns{7}
\setlength{\tabcolsep}{12pt}
\tablewidth{\columnwidth}
\tablehead{
\colhead{Parameter} &
\colhead{Units} &
\multicolumn{2}{c}{Close Solution} &
\multicolumn{2}{c}{Wide Solution} \\
\cline{3-4} \cline{5-6}
 & & \colhead{Value} & \colhead{MCMC Avg} & \colhead{Value} & \colhead{MCMC Avg}
}
\startdata
$t_E$ & days & $55.101$ & $55.147 \pm 0.129$ & $ 55.226$ & $ 55.336 \pm 0.202$ \\
$t_{0}$ & HJD$'$ & $8590.553$ & $8590.555 \pm 0.005$ & $8590.556$ & $8590.556 \pm 0.002$ \\
$u_0$ & {} & $0.057$ & $0.057 \pm 0.001$ & $0.058$ & $0.058 \pm 0.001$ \\
$s$ & {} & $0.921$ & $0.921 \pm 0.005$ & $1.020$ & $ 1.019 \pm 0.002$ \\
$\alpha$ & rad & $-2.102$ & $-2.104 \pm 0.004$ & $-2.105$ & $-2.104 \pm 0.002$ \\
$q (10^{-5})$ & {} & $3.350$ & $3.360 \pm 0.020$ & $4.340$ & $4.320 \pm 0.050$ \\
$t_*{}$ & days & $0.053$ & $0.053 \pm 0.003$ & $0.051$ & $0.051 \pm 0.004$ \\
$\pi_{E, N}$ &  & $0.204$ & $0.203 \pm 0.003$ & $0.203$ & $0.210 \pm 0.015$ \\
$\pi_{E, E}$ &  & $0.074$ & $0.076 \pm 0.006$ & $0.076$ & $0.079 \pm 0.008$ \\
$I_s{}$ & {} & $19.682$ & $19.683 \pm 0.002$ & $19.682$ & $19.684 \pm 0.003$ \\
$V_s{}$ & {} & $22.710$ & $22.711 \pm 0.002$ & $22.710$ & $22.711 \pm 0.002$ \\
\hline
Fit $\chi^2 / \textrm{dof}$ & {} & $23797.19/23620$ & {} & $23801.38/23620$ & {}
\enddata
\end{deluxetable*}


\begin{deluxetable*}{lccc}[!htp]
\deluxetablecaption{Planetary System Properties from Lens Flux Constraints\label{tab:planet-params}}
\tablecolumns{4}
\setlength{\tabcolsep}{17.5pt}
\tablewidth{\columnwidth}
\tablehead{
\colhead{\hspace{-6cm}Parameter} & \colhead{Units} &
\colhead{Values \& RMS} & \colhead{2-$\sigma$ range}
}
\startdata
Angular Einstein Radius ($\theta_E$) & mas & $1.14 \pm 0.03$ & $1.10-1.21$\\
Geocentric lens-source relative proper motion ($\mu_{\textrm{rel,G}}$) & mas/yr & $7.58 \pm 0.20$ & $7.26-8.03$\\
Host mass ($M_{\rm h}$) & $M_{\Sun}$ & $0.65 \pm 0.04$ & $0.58-0.73$\\
Planet mass ($M_{\rm p}$)\\
\hspace{5mm} Close Solution & $M_{\Earth}$ & $7.18 \pm 0.40$ & $6.26-7.94$\\
\hspace{5mm} Wide Solution & $M_{\Earth}$ & $9.48 \pm 1.13$ & $7.41-11.87$\\
2D Separation ($a_{\perp}$) & au & $3.06 \pm 0.12$ & $2.83-3.31$\\
3D Separation ($a_{3\textrm{d}}$) & au & $3.74 \pm 0.15$ & $3.47-4.06$\\
Lens Distance (D$_{L}$) & kpc & $2.64 \pm 0.22$ & $2.22-3.10$\\
Source Distance (D$_{S}$) & kpc & $7.49 \pm 0.53$ & $6.19-8.71$\\
\enddata
\tablenotetext{}{\footnotesize{\textbf{Notes.}} The close and wide solutions yield identical values, with the exception of planet mass ($M_p$).}
\end{deluxetable*}

\begin{figure}
\includegraphics[width= 0.45\textwidth]{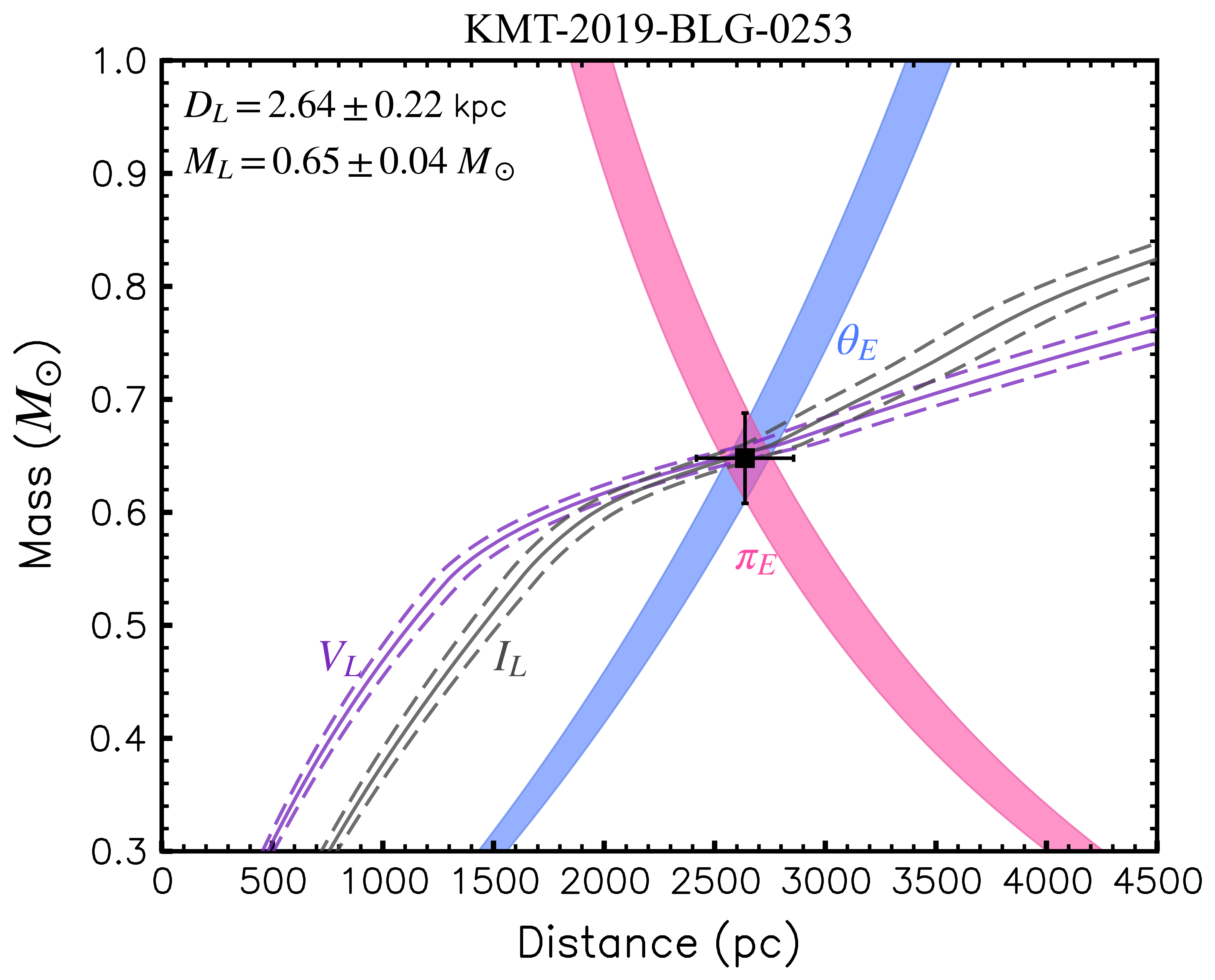} 
\centering
\caption{\label{fig:mass-distance} The mass–distance relation for KB190253 with constraints from the lens flux measurement in HST V (purple) and HST I (grey). Dashed lines show the 1$\sigma$ error bars for each passband. The solid pink region shows the mass–distance relation calculated using the microlensing parallax measurement ($\pi_E$), and the solid blue region shows the mass–distance relation calculated using the angular Einstein radius measurement ($\theta_E$).}
\end{figure}

To measure our new lens system parameters, we sum over the MCMC results using a Galactic model \citep{bennett:2014a} with weights for the microlensing rate and our $\mu_{\textrm{rel,H}}$ value from \textit{HST} (described in Sec. \ref{sec:prop-motion}). Additionally, we include the source distance as a fitting parameter in the re-modeling of the lightcurve with imaging constraints. We include a weighting from the \cite{koshimoto:2021a} Galactic model as a prior for $D_S$, and we also use the same Galactic model to obtain a prior on the lens distance for a given value of $D_S$. This prior is not used directly in the lightcurve modeling, but instead is used to weight the entries in a sum of Markov chain values. These weights affect the posterior distributions of $D_L$ and $D_S$, and consequently propagate through $\pi_{\rm rel}$ into the inferred $\boldsymbol{\mu}_{\rm rel,G}$.
\\
\indent The measurement of the angular Einstein radius allows us to use a mass-distance relation if we assume the distance to the source is known \citep{bennett:2008a, gaudi:2012a}:

\begin{equation}
\label{eqn:ml}
    M_{L} = \frac{c^2}{4G}\theta_{E}\frac{D_{S}D_{L}}{D_{S}-D_{L}},
\end{equation}

\noindent where $M_{L}$ is the lens mass, $G$ and $c$ are the gravitational constant and speed of light. $D_{L}$ and $D_{S}$ are the distance to the lens and source, respectively.\\
\indent Additionally, since we have a direct measurement of lens flux in the $I-$ and $V-$ bands, we utilize the \citet{delfosse:2000a} empirical mass-luminosity relations in each of these passbands. We refer the reader to \cite{bennett:2018a} for a broad description on the use of mass-luminosity relations (both empirical and analytic) for inferring the mass-distance relation for lens systems. We consider the foreground lens extinction in each passband, as described in Sec.~\ref{sec:source-lens} and generate the relations in conjunction with the mass-distance relation given in Equation~\ref{eqn:ml}. Figure~\ref{fig:mass-distance} shows the mass-distance plane with our new direct calculation for the lens mass and distance (black). The red/blue curves represent the constraint from the mass-luminosity relation, with dashed lines representing the error from the \textit{HST} lens flux measurement. Additionally, the $\theta_{E}$ constraint is shown with error primarily from the source distance uncertainty. \\
\indent Table \ref{tab:planet-params} gives the derived lens system physical parameters with RMS errors and 2$\sigma$ ranges. Although we are able to successfully reduce the number of possible binary-lens solutions presented in \citet{hwang:2022a} by a factor of 2, the close/wide degeneracy still remains. Furthermore, we find two significantly different $q$ values for the close and wide solutions. This ultimately affects the mass of the planetary companion for each solution, while the host mass is approximately the same as given by the direct flux to mass-luminosity relationship described earlier. Table \ref{tab:planet-params} reports the lens system physical properties from the wide solution, except for planet mass which we report both close/wide values. Besides the mass ratio $q$, the close and wide solutions largely agree with each other and both give nearly identical best-fit $\chi^2$ values (Table \ref{tab:lcpar_combined}).\\ 
\begin{figure*}
\includegraphics[width= \textwidth]{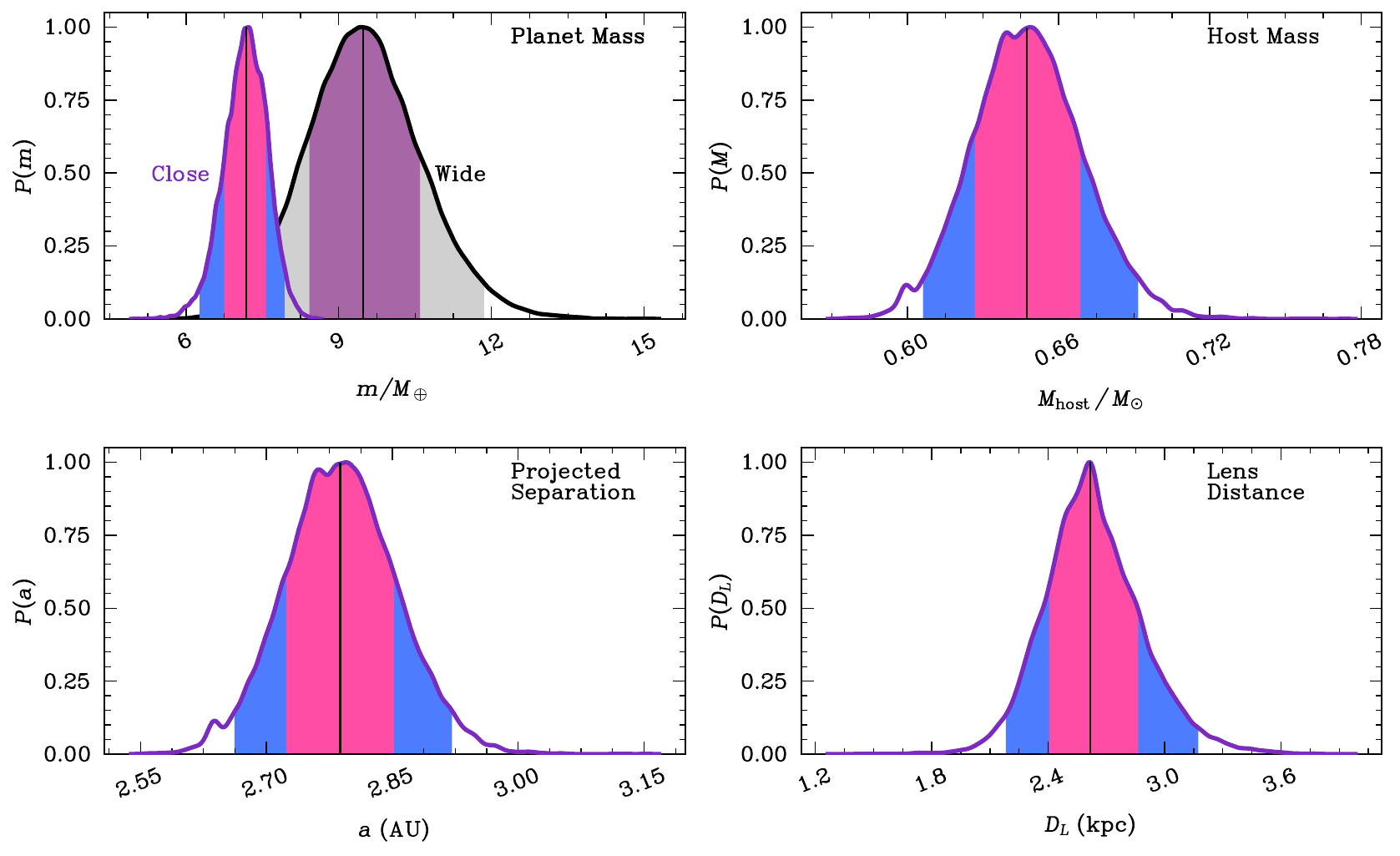} 
\centering
\caption{\label{fig:prob_dist_in} The posterior probability distributions for the lens system physical parameters: planetary companion mass (upper left), host mass (upper right), 2D projected separation (lower left), and lens distance (lower right). In the upper left panel, the blue-and-pink shade represents the close solution, and the purple-and-gray shade represents the wide solution.  The vertical black line shows the median of the probability distribution for each parameter. The central 68\% of each distribution is shown in violet (purple for the wide solution), with the remaining central 95\% shown in dark blue (grey for the wide solution).}
\end{figure*}
\indent We can calculate the planet's semi-major axis from the expression $a_{\perp} = sD_{L}\theta_{E}$, where $s$ is the projected separation given by the lightcurve modeling. We find a 2D separation of $a_{\perp} = 3.06 \pm 0.12$ au. Additionally, we find the lens system is at a distance of $D_L = 2.64 \pm 0.22$ kpc. Further, we find that the lens star has a mass $M_{\textrm{host}} = 0.65 \pm 0.04 M_{\Sun}$, with a planetary companion of mass $m_{\textrm{planet}} = 7.18 \pm 0.40 M_{\Earth}$ for the close solution or $m_{\textrm{planet}} = 9.48  \pm 1.13 M_{\Earth}$ for the wide solution. \\
\indent Using Table~\ref{tab:dual-phot} and under the assumption that Star 2 is the source, we measure $V_S=22.712\pm0.072$ and $I_S=19.685\pm0.049$, corresponding to a color $(V-I)_S=3.027\pm0.087$. Adopting $E(V-I)=2.30$ and $A_I=2.75$ from Sec.~\ref{sec:extinction}, we obtain $(V-I){S,0}=0.727$ and $I_{S,0}=16.935$. We use the surface-brightness relations from \citet{adams2018} to infer an angular source radius of $\theta_* = 1.38 \pm 0.14 \mu as$. Using the close solution values in Table~\ref{tab:lcpar_combined},  $t_*=0.053\pm0.003$ days and $t_E=55.147\pm0.129$ days, we estimate $\theta_E = 1.44 \pm 0.17$ mas. The corresponding microlensing parallax amplitude is $\pi_E= 0.217 \pm0.004$, leading to a lens mass of $M_L = 0.80\pm 0.10 M_\odot$. This value is broadly consistent (within 2 $\sigma$) with the lens mass reported in Table~\ref{tab:planet-params}. \\
\indent The above estimates are consistent with a planet with mass between a super-Earth and sub-Neptune orbiting a K5 to K7 dwarf star. Lastly, in Figure~\ref{fig:prob_dist_in} we show the posterior probability distributions for the planetary companion mass, host star mass, 2D projected separation, and lens system distance.


\section{An automated Pipeline for High-Resolution Image Analysis} 
\label{sec:automated_pipeline}

\begin{figure*}
\includegraphics[width= \textwidth]{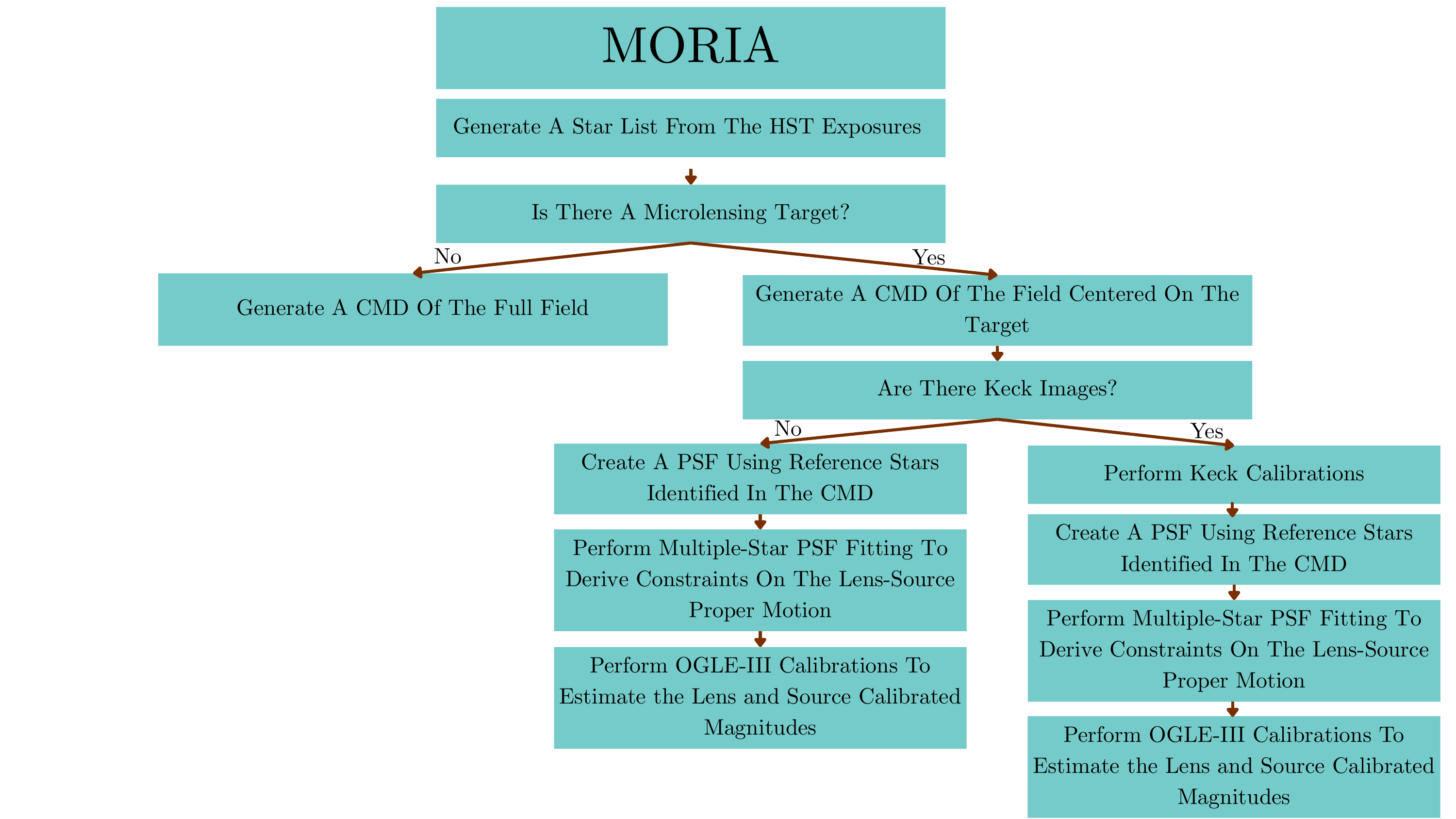} 
\centering
\caption{\label{fig:alg} A generalized flowchart describing how \texttt{MORIA} can be used to perform high-resolution image analysis of targets in \textit{HST}, \textit{Roman}, or other high-resolution observing facilities.}
\end{figure*}

The following section describes the development of the Microlensing Object high-Resolution Imaging Analysis \MORIA software. \MORIA is a Python package designed to automate the high-resolution image analysis and PSF fitting process detailed first by \citet{bennett:2015a}. This is the first major component of the higher-level image-constrained lightcurve modeling procedure. \MORIA was used in this study to derive the OGLE-calibrated HST magnitudes presented in Table~\ref{tab:dual-phot}, generate the multiple-star PSF fitting seen in Figure~\ref{fig:4panel_hst} and to obtain the separation (and $\boldsymbol{\mu}_{\rm rel}$) constraints shown in Table~\ref{tab:combined_sep}. \MORIA is fully open-source, available on GitHub\footnote{\url{https://github.com/10ay/MORIA}}, and pip-installable. Documentation of each major module in the code, as well as a detailed description of how to run \MORIA is provided in various Python notebooks via the repository. \\
\indent \MORIA has several dependencies that are required to be installed before running; these include typical libraries like NumPy \citep{oliphant:2006a}, Astropy \citep{robitaille:2013a}, pandas \citep{reback2020pandas} and matplotlib \citep{hunter:2007a}. Additionally, \texttt{MORIA}'s backend involves running \texttt{Fortran} codes, so a \texttt{Fortran} compiler is required. Currently, upon installation, a ``Makefile.txt'' file runs in the \texttt{Fortran\_compile} folder. The ``Makefile.txt'' uses \texttt{gfortran} as the default \texttt{Fortran} compiler during installation. If a user wishes to use a different compiler, the ``Makefile.txt'' must be edited accordingly.\\
\indent The primary functionality of \MORIA, as it pertains to HST-observed microlensing targets, has been summarized in Figure~\ref{fig:alg}. \\
\indent In Sec.~\ref{sec:cmd}, we explain the first major steps of \texttt{MORIA}; generating star lists and color-magnitude diagrams from the high-resolution images. In Sec.~\ref{sec:psf_create}, we describe the process of star selection to create an empirical point-spread function (PSF) model with similar properties as the target. In Sec.~\ref{sec:psf_fit}, we explain how the pixels around the target are fit with the PSF model to determine the best-fit two or three star solution. We detail the photometric calibration procedure in Sec.~\ref{sec:ogle}. This step is used to place the \textit{HST} instrumental magnitudes into the OGLE-III photometric system for the $V$ and $I$ passbands. In some cases there are high-resolution images available from both \textit{HST} and Keck for a given target. In Sec.~\ref{sec:keck} we describe an optional function in \MORIA that allows for astrometric calibration between Keck and \textit{HST} reference frames. 

\subsection{Data Reduction and Color-Magnitude Diagram\label{sec:cmd}}

The initial input to \MORIA must be \textit{HST} imaging data from the Wide-Field Camera 3 (WFC3) instrument. For our study, we used data from program GO-17834 \citep{terry:2024prop} which took images of the microlensing target KB190253 in the $F814W$ and $F606W$ passbands. We obtained the calibrated \texttt{\_flc} exposures from the Mikulski Archive for Space Telescopes (MAST). The \texttt{\_flc} files are the standard \textit{HST} pipeline-processed images for many cameras onboard the space telescope.  These data include corrections for the charge-transfer efficiency (CTE) degradation which naturally occurs over time as high-energy particles impact the detectors during their lifetimes. \MORIA first converts these \texttt{\_flc} exposures into \texttt{\_WJ2} files. The \texttt{\_WJ2} files store the calibrated images in a common full-chip coordinate system and apply the geometric distortion corrections required for subsequent astrometric transformations and PSF analysis.\\
\indent A suite of library PSF models from \cite{anderson:2006a} were initially used to measure stellar positions and fluxes in each exposure. This step produces \texttt{.xym} files, which are standard star-list files containing the measured stellar positions, instrumental magnitudes, and fit quality metrics for every detected source in an image. The library PSFs are based on spatially varying effective PSF models, which can be further perturbed by measuring local distortions in a moving window across the grid of pixels. \\
\indent Individual star lists are then transformed into a common reference frame that accounts for geometric distortions inherent in the HST detectors. An iterative procedure is used to get a list of photometry that allows small zeropoint shifts for each exposure. The final transformation step yields a stacked image and a ``matchup'' file containing the averaged photometry, astrometry and related errors for all detected stars.\\
\indent Using the matchup files, an instrumental CMD can be generated. The CMD is used to identify several important stars with colors and magnitudes similar to the target that will be used to construct a PSF model with similar characteristics as the target. The matchup files also select brighter stars near the target that can be used to refine the coordinate transformation and that can be used to calibrate the HST photometry to the OGLE-III catalog later in the process. Note that the output of this step does not create a figure analogous to Figure~\ref{fig:CMD}. This step in MORIA occurs prior to OGLE-III photometric calibrations. The OGLE-III calibrations are performed at a later stage, as described in Section~\ref{sec:ogle}.

\subsection{Creating a Point-Spread Function Model \label{sec:psf_create}}
Using similar-brightness nearby reference stars identified in the image and CMD, local coordinate transformations are derived for each exposure to account for small residual geometric distortions and frame-to-frame offsets. Pixel-level values surrounding the target star and the selected PSF-star candidates are extracted from each exposure and transformed into a reference frame corrected for distortion. The extracted pixel list contains the flux of each pixel relative to the star centers, this is fundamental for building the empirical PSF model. Reference star selection based on locations on the CMD is of particular importance, as this can help to minimize color-dependent effects on the PSF.\\
\indent For each filter, using nearby stars selected to have similar magnitudes and colors to the target star, \MORIA builds a PSF model. Candidate PSF stars are evaluated using residual images from PSF subtraction and stars that show significant residuals due to blending or an excess of Poisson noise are excluded in a second iteration used for PSF modeling. The final output of this step is a high-resolution, empirically-derived PSF model for each filter.

\subsection{Fitting a Point-Spread Function Model \label{sec:psf_fit}}
The PSF model is then used to fit the pixels surrounding the target star with one-star, two-star or three-star models. The fitting of multiple PSF models occurs simultaneously, as opposed to iteratively. The simultaneous nature of the fitting preserves correlations between separations and flux ratios that may otherwise be missed when fitting in an iterative manner.\\
\indent As described in Sec.~\ref{sec:multi-star}, separate instances of one-star, two-star, and three-star fitting were conducted on the pixel grid for KB190253. For a given model, \MORIA constructs a forward model of the observed pixel grid by shifting and scaling the empirical PSF template to the trial positions and fluxes of all stellar components. The free parameters include the relative star positions and the individual star fluxes.\\
\indent Parameter estimation is performed using Markov Chain Monte Carlo (MCMC) sampling. At each MCMC step, the model PSFs placed at a position on the pixel grid and given a flux value. The MCMC sampling performs a standard Metropolis-Hastings method to accept or reject proposed models via the $\chi^2$ evaluation. Each accepted or rejected model, as well as the global best-fit solution for stellar positions and fluxes are saved to output files. Outlier pixels with a $\chi^2_{\textrm{pix}}$ value above a user-defined threshold are rejected during this process.\\
\indent For KB190253, the PSF fitting of one, two and three-stars are displayed in the top-right, bottom-left and bottom-right panels of Figure~\ref{fig:4panel_hst}. By default \MORIA performs one- and two-star fitting. However, an argument can be passed to run the three-star fit when necessary. Simultaneous fitting of $n > 3$ objects is not currently supported in \MORIA (see future improvements in Sec.~\ref{sec:future_moria}).

\subsection{OGLE-III Calibrations \label{sec:ogle}}

As a final step, \MORIA can be used to calibrate the HST photometry to the OGLE-III photometric system. Selected calibration stars in HST are remeasured using the final PSF models, and these measurements are matched to the OGLE-III catalog. Special efforts are required to avoid blending effects, where several HST stars will be merged together to make an object that is considered to be a single star in the lower angular resolution OGLE images. Photometric offsets are then derived to convert instrumental HST magnitudes into calibrated OGLE-III-equivalent magnitudes.  \\
\indent Using the identified target coordinates, \MORIA directly obtains the OGLE-III field number, chip number, and pixel coordinates from the OGLE database using the ``OGLE Field Finder” tool within the ``Sky Coverage” page. Next \MORIA chooses the appropriate OGLE-III catalog file and downloads the corresponding photometry map and FITS image for a given target. In the case of KB190253, the OGLE-III photometry catalog file is \texttt{blg194.1.map} and the corresponding FITS image file is \texttt{blg194.I.1.fts}.\\
\indent We note part of this step in \MORIA is not automated. To fully calibrate the HST image with OGLE-III, coordinates must be matched by eye between the OGLE reference image and the HST image. Once stars are identified, their precise coordinates can be determined by running \texttt{star2reg.py} directly on the OGLE-III catalog. \texttt{star2reg.py} creates a star list map that can be overlaid on top of the HST image. This helps alleviate some of the difficulty in physical eye-matching of coordinates between the two frames. This assists the user in visually identifying common stars between the OGLE and HST reference frames, even when the images often have different orientations and angular resolutions. Once the common stars have been cross-identified, the final photometric calibration step with \MORIA can be run. The results of OGLE-III calibrations are routed to ``.log" files. These calibrations include the OGLE-III calibrated magnitudes (Table~\ref{tab:dual-phot} for KB190253) that can be used to create the calibrated CMD diagram seen in Figure~\ref{fig:CMD}.

\subsection{Optional Keck Calibrations \label{sec:keck}}
\MORIA has the ability to (optionally) incorporate Keck images of a given target to derive an astrometric transformation between Keck and \textit{HST}. KB190253 does not currently have Keck observations, therefore this step is skipped in our current analysis. Running the Keck calibrations in \MORIA requires two files from Keck: a NIRC2-narrow camera or OSIRIS image, and a Keck star list. For targets with Keck observations, we typically use the Keck AO Imaging (KAI) pipeline \citep{lu:2022a} and \texttt{DAOPHOT} \citep{stetson:1987a} to generate the stacked image and star list files that are required by \MORIA. \\
\indent A user is required to identify a set of relatively isolated, bright (but not saturated) stars in both the Keck and HST images. Once this is done, \MORIA proceeds to fit for the coordinate transformation from Keck to HST coordinates. This is necessary if one wishes to impose constraints on the lens-source separation in the HST images based on the Keck analysis. This procedure automatically removes outliers and yields an RMS scatter in each of the x and y directions of about 0.025 HST pixels or about 1 mas (assuming well-sampled HST data and good-quality Keck data).

\section{Applications of \MORIA to Roman \label{sec:roman_moria}}


The upcoming \textit{Roman} GBTDS will produce high-resolution images for thousands of microlensing targets \citep{penny:2019a, terry:2026a}. Lens flux constraints from Roman's images will help rectify possible systematic errors, and could be used to break microlensing degeneracies that can arise from ground-based (seeing-limited) microlensing modeling \citep{bennett:2024a}.\\
\indent \MORIA has been developed as a framework to demonstrate the utility of such an automated pipeline, capable of extracting lens and source properties from crowded-field images, and using these direct measurements to place high-precision constraints on lens system physical properties (Sec.~\ref{sec:lens-properties}). This can enable quick characterization of lens and source stars across the very large expected sample of events from the mission. With appropriate modifications to accommodate the characteristics of the \textit{Roman} Wide-Field Instrument (WFI), \MORIA can be adapted as a pipeline for \textit{Roman} microlensing target analysis. 

\subsection{KMT-2019-BLG-0253 in Roman GBTDS Data} \label{sec:kb190253_gbtds}
The target KB190253 is located at Galactic coordinates ($\ell,b=(0.126, -1.429)$), which places it inside the expected \textit{Roman} GBTDS footprint\footnote{\url{https://roman.gsfc.nasa.gov/science/ccs/ROTAC-Report-20250424-v1.pdf}}. This means that the additional epoch of high-resolution followup imaging mentioned earlier can be conducted using well-sampled GBTDS imaging data. Using our measured $\mu_{\textrm{rel,H}}$ value from \textit{HST} (Table \ref{tab:combined_sep}), we estimate that the lens and source will be separated by approximately 65 mas during the first Season of the GBTDS (between February and April 2027). Although the \textit{Roman} WFI pixels are larger than \textit{HST} WFC3-UVIS pixels, \textit{Roman} will obtain thousands of exposures of the target (compared to 16 for \textit{HST}). This will allow for exquisite sampling of the \textit{Roman} pixel phase.\\
\indent The lens, source, and blend are of similar brightness in the $V$ and $I-$bands (Table \ref{tab:dual-phot}). If we assume they are of similar brightness ($\pm 1$ mag) in the \textit{Roman} $F146$ very wide passband, then all three targets should be easily detectable with a \texttt{MORIA}-like analysis in the Season 1 \textit{Roman} data. As mentioned in Sec. \ref{sec:source-lens}, obtaining an additional epoch of high-resolution imaging (i.e. \textit{Roman} GBTDS) will help to definitively identify the source, lens, and the nature of the blend (third) star. This future measurement will confirm or reject our initial identifications made in this work.

\subsection{Future Improvements to \texttt{MORIA}}\label{sec:future_moria}
We have demonstrated in this work \MORIA can be used as a powerful tool for precise determination of source and lens properties in microlensing targets. We discuss here potential future development and improvements to the structure of \MORIA that are currently beyond the scope of the current work. A list of improvements and future developments may include:

\begin{enumerate}
    \item Support for \MORIA analysis on \textit{HST}-ACS images, \textit{JWST} images, and future \textit{Roman} images.
    \item Added functionality for simultaneous PSF fitting of $n > 3$ objects.
    \item Parallelization and/or batch processing of large numbers of targets, as expected to be delivered by the \textit{Roman} GBTDS mission.
    \item A \MORIA hook to external software developed for \textit{Roman/HST/Euclid} joint analyses of GBTDS microlensing events (e.g. \texttt{HAMRR}; \cite{terry:2026b}).
\end{enumerate}



\section{Discussion and Conclusion} \label{sec:conclusion}
Image-constrained modeling will be critical for the interpretation of planetary microlensing systems discovered by the upcoming RGES survey. RGES will use the same methods to determine the masses of and distance to planetary microlensing systems as we have demonstrated with high angular resolution followup analysis via \textit{HST}. There is a need to develop an automated high-resolution image analysis pipeline in preparation for the tens of thousands of microlensing events that RGES will measure. In this paper, we have presented \MORIA as an automated pipeline to go from high-angular resolution images to precise constraints on lens-source separations, relative proper motions, and brightnesses.\\
\indent We developed and tested \MORIA on high-resolution followup observations of the microlensing target KMT-2019-BLG-0253. We have made a direct measurement of flux from the host star as well as a precise determination of the direction and amplitude of the lens-source relative proper motion. Using the multiple-star PSF fitting routine in \MORIA, we find strong evidence for three blended stars that comprise the target. Compared to the original study of KMT-2019-BLG-0253 \citep{hwang:2022a}, the uncertainty on the parallax $\pi_E$ value we determined is significantly smaller and tightly constrained. While \citet{hwang:2022a} placed a lower-bound on $\theta_E$ and $\boldsymbol{\mu}_{\rm rel}$, our re-analysis of the microlensing lightcurve photometry that includes constraints from the high-resolution \textit{HST} detections enables us to measure these values with significantly higher precision (Table \ref{tab:planet-params}).\\ 
\indent The lens flux measurements we make enable us to use mass-luminosity relations and new constraints on higher-order lightcurve effects ($\pi_E$ and $\theta_E$) to measure a precise mass and distance for the lens system. The KMT-2019-BLG-0253 microlensing event has a planet-to-star mass ratio of $q = 3.36\pm 0.02\times 10^{-5}$ for the close solution and $q = 4.32\pm 0.03\times 10^{-5}$ for the wide solution. We find the lens system distance to be $D_L = 2.64 \pm 0.22$ kpc, which is somewhat smaller than the $4.9^{+1.9}_{-1.6}$ kpc estimated by \citet{hwang:2022a} who used a Bayesian analysis. Lastly, we estimate the projected star-planet separation to be $ a_{\textrm{3d}} = 3.74 \pm 0.15$ au.\\
\indent We have re-modeled the lightcurve photometry using the image-constrained modeling version of \texttt{eesunhong} \citep{bennett:1996a, bennett:2010a}. The image-constrained modeling has allowed us to incorporate all available lightcurve observables and high-resolution imaging observables in a holistic manner to reduce the number of possible solutions by a factor of two and achieve precise physical parameters for the lens system. The classical close/wide degeneracy remains for this event. \\
\indent Finally, the results of this work have several implications for the upcoming \textit{Roman} GBTDS and in particular the RGES microlensing survey. Using a framework like \MORIA to perform high-resolution image analysis with \textit{Roman} data will be broadly beneficial to the \textit{Roman} microlensing community.

\section*{Acknowledgements}
\noindent Support for program GO-17834 was provided by NASA through a grant from STScI, which is operated by AURA, Inc., under NASA contract NAS 5-26555. TDB, SKT, DPB, AB, JH, and AV are supported by the Nancy Grace Roman Space Telescope Project through the National Aeronautics and Space Administration grant 80NSSC24M0022. The HST data presented in this article were obtained from the Mikulski Archive for Space Telescopes (MAST) at the Space Telescope Science Institute. The specific observations analyzed can be accessed via \dataset[doi: 10.17909/hvam-1n56]{https://doi.org/10.17909/hvam-1n56}.

\textit{Software}: Astropy \citep{astropy:2013, astropy:2018, astropy:2022}, eesunhong \citep{bennett:1996a, bennett:2024a}, genulens \citep{koshimoto:2021a}, Matplotlib \citep{hunter:2007a}, MORIA (this work), Pandas \citep{reback2020pandas}, Numpy \citep{harris:2020array}

\bibliographystyle{aasjournal}
\bibliography{kb190253.bib}

\end{document}